\documentclass[12pt,preprint]{aastex}

\newcommand{\Chandra}{{\it Chandra}}
\newcommand{\XMM}{{\it XMM-Newton}}
\newcommand{\ROSAT}{{\it ROSAT}}
\newcommand{\ASCA}{{\it ASCA}}
\newcommand{\cdegree}{$^\circ$C}

\newcommand{\solarM}{$\rm M_{\odot}$}
\newcommand{\solarZ}{$\rm Z_{\odot}$}
\newcommand{\solarLB}{$L_{\rm B,\odot}$}
\newcommand{\apec}{APEC}

\slugcomment{}

\begin{document}

\shorttitle{Abundance Distribution in HCG 62}
\shortauthors{Gu et al.}

\title{A High-Abundance Arc in the Compact Group of Galaxies HCG 62:
An AGN- or Merger-Induced Metal Outflow?}
\author{Junhua Gu\altaffilmark{1},
Haiguang Xu\altaffilmark{1},
Liyi Gu\altaffilmark{1},
Tao An\altaffilmark{2},
Yu Wang\altaffilmark{1},
Zhongli Zhang\altaffilmark{1},
and Xiang-Ping Wu\altaffilmark{3}}

\altaffiltext{1}{Department of Physics, Shanghai Jiao Tong
University, 800 Dongchuan Road, Shanghai 200240, PRC;
e-mail:
\mbox{tompkins}@sjtu.edu.cn, hgxu@sjtu.edu.cn,
\mbox{alfred\_gly}@sjtu.edu.cn,
\mbox{wenyu\_wang}@sjtu.edu.cn,
\mbox{zebrafish}@sjtu.edu.cn}

\altaffiltext{2}{Shanghai Astronomical Observatory, Chinese Academy of Sciences,
80 Nandan Road, Shanghai 200030, PRC;
e-mail: antao@shao.ac.cn}
\altaffiltext{3}{National Astronomical Observatories, Chinese
Academy of Sciences, 20A Datun Road, Beijing 100012, PRC;
e-mail: wxp@bao.ac.cn}

\begin{abstract}
We present a \Chandra\ study of the metal distribution in the X-ray bright compact
group of galaxies HCG 62. We find that the diffuse X-ray emission is peaked at the
core of the central galaxy NGC 4778, and is dominated by the contribution of the
hot gas. The diffuse emission is roughly symmetric within $\simeq0.25^{\prime}$,
which is straddled by double-sided X-ray cavities aligned in the northeast-southwest
direction. By mapping the emission hardness ratio distributions and by performing
the 2-dimensional spectral analysis, we identify a remarkable high-abundance arc
region at about
$2^{\prime}$ ($33.6h_{70}^{-1}$ kpc)
from the X-ray peak that spans over a vast region from south to northwest, a part
of which roughly coinciding with the outer edge of the southwest X-ray cavity. The
measured average abundance in this arc is higher than that in its neighboring regions
by a factor of about 2, and the abundance ratios therein are nicely consistent with
the dominance of the SN Ia yields. We estimate that the mass of iron contained in
the arc is
$>3\times 10^{6}h_{70}^{-2.5}$ \solarM,
which accounts for $>3$\% of the iron synthesized in the galaxy. The high-abundance
arc could have been formed by the AGN activities. However, it is also possible that
the arc was formed in a recent merger as is implied by the recent optical kinematic
study (Spavone et al.\ 2006), which implies that mergers may be as important as AGN
activities in metal redistributions in early-type galaxies and their associated groups
or clusters.
\end{abstract}


\section{Introduction}
Measurements of the metal distribution in hot plasma in clusters and groups of
galaxies can provide us with important constraints on the physics of feedback
processes, which is tightly related to the study of supernova rates, stellar
winds and gas heating on the group/cluster scales. Early X-ray observations of
E/S0 galaxies showed that the iron abundance of the hot interstellar medium (ISM)
is sub-solar and is significantly lower than the theoretical predictions based
on the standard galactic wind and supernova synthesis models (e.g.,
Arimoto et al.\ 1997). In recent years this problem has been mitigated to some
extent by the observations with \Chandra\ and \XMM\, which show clear evidence
that the gas abundance is close to or slightly over the solar abundance in the
inner regions of bright E/S0 galaxies (e.g., Buote 2002; Xu et al.\ 2002;
Matsushita et al.\ 2003; Humphrey and Buote 2006), although the sub-solar abundance
still cannot be ruled out unambiguously for the X-ray faint systems
(e.g., Sarazin et al.\ 2001; Irwin et al.\ 2002; O'Sullivan \& Ponman 2004). It
also has been revealed that the iron abundance distributions in a few bright
clusters of galaxies are not simply central-peaked, which include the Perseus
Cluster (Sanders et al.\ 2005), Abell 1060 (Hayakawa et al.\ 2004, 2006), Abell
2199 (Johnstone et al.\ 2002) and AWM 7 (Furusho et al.\ 2003). In these clusters
the highest abundances are observed in off-center regions located at from a few
kpc to a few tens kpc from the cluster center. It has been suggested that the
observed fluctuations in the abundance distributions may be caused by metal-enriched
bubbles lifting from the cluster core (Sanders et al.\ 2005; Mathews et al.\ 2003),
resonance scattering (e.g., Gil'fanov et al.\ 1987), or inhomogeneous cooling
in the thermal evolution history (Brighenti \& Mathews 2005). In the Perseus
Cluster the high-abundance region was reported to coincides with the edge of
a radio mini-halo, indicating that it could have been formed by the AGN
activities in the recent past.

In this work, we present a \Chandra\ study of the 2-dimensional metal
distribution in the hot gas of the compact group of galaxies HCG 62 ($z=0.0137$;
Hickson et al.\ 1992). HCG 62 is a gas-rich, luminous group in X-rays
(Lx=$1.10\times10^{43}$ erg s$^{-1}$; Ponman et al.\ 1996). It contains 4
bright member galaxies and is dominated by the E3/S0 galaxy NGC 4778, the
brightest member in both optics and X-rays (Pildis et al.\ 1995) that is
identified as a low-luminosity AGN (Coziol et al.\ 1998). Kinematic study
of the rotation curve and velocity dispersion indicated that NGC 4778 is
possibly interacting with another member NGC 4761 (Spavone et al.\ 2006).
One other luminous member NGC 4776 (S0) is located at a projected distance
of only $\simeq0.4^{\prime}$ west of NGC 4778 with a peculiar velocity of
737 km s$^{-1}$. However, it shows no signs of interaction or merging with
NGC 4778 (Rampazzo et al.\ 1998). In this paper, we adopted a distance of
57.7 Mpc to the group by using $H_{0}=70h_{70}^{-1}$ km s$^{-1}$ Mpc$^{-1}$,
$\Omega_{m} = 0.3$ and $\Omega_{\Lambda} = 0.7$. We utilized the solar
abundance standards of Grevesse and Sauval (1998), where the iron abundance
relative to hydrogen is $3.16\times10^{-5}$ in number.

\section{Observation and Data Reduction}
The \Chandra\ observation of HCG 62 was carried out on January 25, 2000 (ObsID 921)
for a total exposure of 49.2 ks with CCDs 3, 5, 6, 7 and 8 of the Advanced CCD
Imaging Spectrometer (ACIS) in operation. The center of the group-dominating galaxy
NGC 4778 was positioned close to the normal aim point on the S3 chip (CCD 7) with
an offset of $2.3^{\prime}$. The events were collected with a frame time of 3.2 s
and telemetried in the Faint mode. The focal plane temperature was set to -110
\cdegree. Since for the time being 1) the effects of the ACIS CTI and its corrections
on the abundance measurements are not fully clear, 2) for the iron abundances
below 1.5 solar the non-CTI-corrected and CTI-corrected results show only little,
negligible difference between them, and 3) in general, the results of the spectral
analysis with the non-CTI-corrected ACIS data are nicely consistent with those
obtained with the \XMM\ data, with the oxygen abundance being the only notable
exception (Sanders \& Fabian 2006a), in this work we used the standard \Chandra\
data analysis package CIAO software (version 3.0) and applied the latest CALDB
(version 3.2.3) to process the data extracted from the S3 chip without applying
any CTI correction.
We kept events with \ASCA\ grades 0, 2, 3, 4 and 6, and removed all the bad pixels,
bad columns, columns adjacent to bad columns and node boundaries. We examined the
lightcurves of the source-free regions on the backside-illuminated S1 (CCD 5) and
S3 chips and detected no strong occasional background flares. To estimate the
contribution of the diffuse emission in the S3 boundary regions (\S3.2.1) we
analyzed the \ROSAT\ PSPC data of HCG 62 acquired on December 27, 1995, which
lasted for 19.6 ks. We followed the standard \ROSAT\ PSPC data analysis procedures
to process the data by using XSELECT (version 2.3) and FTOOLS (version 6.0.3). We
also analyzed the VLA 20 cm (L-band) observations of HCG 62 obtained on June 16,
2001 by using the Astronomical Image Processing System (AIPS) of NRAO following
the standard procedures. Absolute flux density was calibrated by the use of the
3C 48 data, whose flux density was set to be 15 Jy at the selected wavelength.
The calibration error is less than 5\% in the VLA data reduction.

\section{X-Ray Imaging Spectroscopy}
\subsection{Morphology and Surface Brightness Profiles}
In Figure 1a we show the smoothed 0.7--7 keV \Chandra\ ACIS S3 image of HCG 62 in
square-root scale, which has been corrected for exposure but not for background.
We find that the position of the X-ray peak (RA=12h53m05.7s Dec=-09d12m15.2s J2000)
is consistent with the optical center of the dominating galaxy NGC 4778
(RA=12h53m05.9s Dec=-09d12m16.3s J2000; Hickson et al. 1989) to within about
$1.5^{\prime\prime}$. The strong diffuse X-ray emission covers the whole S3 chip,
and is roughly symmetric within the central $\simeq0.25^{\prime}$ region that is
straddled by double-sided X-ray cavities aligned in the northeast-southwest
direction. The two cavities have similar sizes with a projected linear extent
of about $0.6^{\prime}$ ($10.1h_{70}^{-1}$ kpc) for each. Except for the cavities
no other apparent X-ray substructure is identified. To study the X-ray morphology
more quantitatively we extracted the counts in four strip regions (Fig.\ 1b) and
plotted the exposure-corrected count distributions in Figure 1c. We find that in
about $1.2^{\prime}-2^{\prime}$ from the X-ray peak the counts are higher in
southwest than in other directions at the 90\% confidence level. We will discuss
the possible origin of this surface brightness excess in \S4.

About 20 point sources can be resolved visually on the \Chandra\ image, of which
18 sources are detected at the confidence level of $3\sigma$ with the CIAO tool
celldetect and wavdetect. One of them is detected at the center of NGC 4776 to
within 0.5\arcsec, which is possibly a low-luminosity AGN hosted in NGC 4776, and
one other is identified with a foreground star (HD 111960; Hog et al. 1998). No
optical or radio counterpart of the rest detected X-ray point sources can be found
in available literatures.

In Figure 1d we show the 1.43 GHz intensity contour map of HCG 62. The core of
HCG 62 is identified as an extended, weak radio source that has a flux density of
$4.2\pm0.21$ mJy/b. The emission distribution is roughly round within about
0.3$^{\prime}$, which extends farther towards the south so as to form a jet-like
structure that tends to link to a weak radio component located at about
$0.82^{\prime}$ south of NGC 4778. We notice that the jet-like structure coincides
with one of the X-ray cavity exactly, suggesting that the cavities may be fossil
radio bubbles. This supports the conclusion of Coziol et al.\ (1998) that NGC 4778
is an AGN in relatively quiescent state, although we failed to detect the central
point source in either the X-ray band or the radio band at a significance of 3$\sigma$
(see also Morita et al. 2006).

\subsection{Spectral Properties and Abundance Distribution}
\subsubsection{Background}
By studying the \ROSAT\ PSPC surface brightness profile in $0.2-2$ keV we find
that the diffuse X-ray emission of HCG 62 extends outwards to at least $10^\prime$
where the count rate is about $3\sigma$ above the mean background value. To construct
the local \Chandra\ S3 background spectrum, we extracted the spectrum of a
$2.89^{\prime} \times0.22^{\prime}$
region at the boundary of the S3 chip that is adjacent to the S4 (CCD 8) chip, and
then subtracted the contaminating gas component from it. We modeled the gas emission
with an absorbed \apec\ component, whose temperature ($kT$=1.4 keV) and abundance
($Z$=0.45 $Z_\odot$) are obtained from the \ROSAT\ PSPC spectral analysis for roughly
the same region. We find that the obtained S3 background has a spectral energy
distribution consistent with that of the source-free region on the S1 chip and that
used in Morita et al. (2006). In 0.2--2 keV, where the particle and instrument
components are minor, the count rate of the obtained S3 background
($3.5\times10^{-3}$ cts s$^{-1}$ arcmin$^{-2}$) is close to the average X-ray
background count rate calculated from the \ROSAT\ All-Sky Survey diffuse background
maps ($2.5\times10^{-3}$ cts s$^{-1}$ arcmin$^{-2}$). We also have crosschecked our
spectral fittings either by choosing background regions with different boundary
locations or by applying the \Chandra\ blank field spectra alternatively, and
obtained essentially the same best-fit spectral parameters. This is not surprising
because our source spectra were extracted in relative inner ($<3^\prime$) regions
where the X-ray background is overwhelmed by the emission of the galaxy in the
selected bandpass.

\subsubsection{Azimuthally Averaged Analysis}
After removing all the detected point sources we first performed both the projected
and deprojected analysis of the \Chandra\ spectra extracted in 6 annular regions,
which span over $0^{\prime}-2.65^{\prime}$ and are all centered on the X-ray peak.
We limited the fittings to 0.7--7 keV to minimize the effects of the instrumental
background at higher energies and the calibration uncertainties at lower energies.
To compensate for the degradation of the ACIS energy resolution we included an
additional 5\% systematic error in the spectral fittings. We fitted the spectra by
using a model that consists of an APEC (gas) and a power-law (unresolved low-mass
X-ray binaries and background sources) component, both subjected to a common
absorption. In the deprojected analysis, we used the XSPEC model \emph{projct} to
evaluate the influence of the outer spherical shells onto the inner ones. Since
allowing the absorption to vary did not improve the fit, we fixed it to the Galactic
value $N_{\rm H}=3.00\times10^{20}\mbox{cm}^{-2}$ (Dickey \& Lockman, 1990). We
find that the contribution of the hard component is generally less than 10\% of
the total flux, if the photon index is fixed to 1.7, the typical value for the
point source population in early-type galaxies. The fittings are marginally
acceptable or poor for the two innermost annuli in the projected analysis, while
for all the annuli the fittings are acceptable in the deprojected analysis
(Table 1). The best-fit deprojected gas temperature increases steadily from about
$\simeq0.7$ keV at the X-ray peak to $>1$ keV in $^{>}_{\sim}1.2^{\prime}$,
suggesting the group origin of gas in the outer regions. In both the projected
and deprojected analysis the obtained average abundances are nearly consistent
with a constant, except that in $0^{\prime}-0.44^{\prime}$ ($0-7.4h_{70}^{-1}$ kpc)
and $1.77^{\prime}-2.21^{\prime}$ ($29.7-37.1h_{70}^{-1}$ kpc) the abundances
are significantly higher at the 90\% confidence level; the deprojected abundances
of these two regions are $0.86_{-0.27}^{+0.36}$ \solarZ\ and $1.42_{-0.49}^{+1.24}$ \solarZ,
respectively. We notice that in general our results agrees nicely with those
obtained with \Chandra\ and \XMM\ by Morita et al. (2006) at the 90\% confidence
level.

\subsubsection{2-Dimensional Gas Temperature and Abundance Distributions}
In order to get more insight into the abundance gradient we first plot the
2-dimensional distribution of the hardness ratio of the emission in $0.85-1.2$ keV,
a significant part of which is from the blended Fe-L lines, to the continuum
emission extracted in 0.3-0.65 keV and 1.4-5 keV (Fig.\ 2). We find that at
the 90\% confidence level the hardness ratio distribution is highly asymmetric;
the relative iron emissions is significantly enhanced in two regions at about
$2^{\prime}$ southwest and northeast of the X-ray peak, respectively. The two
regions are linked together by an arc-like region showing high hardness ratios,
a part of which roughly coincides with the outer edge of the southwest X-ray
cavity with the jet-like radio structure pointing towards it. We notice that
the arc structure approximately lies within the $1.77^{\prime}-2.21^{\prime}$
annulus where a high abundance value is found in the azimuthally averaged
analysis. In the central region where the average abundance is found high
as well, the iron line emissions also shows remarkable, spatially asymmetric
dominance over the continuum emissions.

Since the high iron line-to-continuum emission ratios can be ascribed to variations
in either gas abundance, or temperature, or both of them, we then probed the
2-dimensional distributions of the gas temperature and abundance simultaneously
via direct spectral modelings utilizing both the Centroidal Voronoi Tessellation
(CVT) algorithm with a conservative threshold of 800 photon counts per cell and
the pie-binning algorithm. Since the contribution of the hard spectral component
is expected to be rather weak, we fitted the $0.7-7$ keV spectra extracted in
each CVT and pie cell by using a single absorbed APEC model with the absorption
fixed to the Galactic value; allowing the absorption to vary or choosing an
absorbed APEC + power-law model does not improve the fittings or change the
best-fits significantly. When it is necessary we used the gehrels weight for
small number statistics, which is defined as $w=1+\sqrt{0.75+N}$ (Gehrels 1986).
The obtained temperature and abundance maps are shown in Figure 3. To examine
if these results are biased by the degeneracy of temperature and abundance
in the spectral fittings, on the temperature maps we also overlaid the contours
of the exposure-corrected hardness ratio of the emission in 1.4--5 keV to that
in 0.3--0.65 keV, which are both continuum-dominated and thus can be used to
trace the temperature gradients. We find that the obtained temperature distribution
in the CVT bins is consistent with that obtained in the pie bins, both approximately
following distribution of the continuum hardness ratio contours. The temperature
maps show that the 2-dimensional temperature distribution in the group is roughly
symmetric, exhibiting a relatively cold core and a tendency for slightly higher
values at $^{<}_{\sim}1^{\prime}$ in the south and west of NGC 4778.

The 2-dimensional abundance maps obtained with the two binning methods agree
with each other as well. At the 90\% confidence level we find that there is an
asymmetric high-abundance arc region, which spans over about 120 degrees azimuthally
from south to northwest of the X-ray peak and roughly coincides with the outer
edge of southwest X-ray cavity in part. The high-abundance arc approximately
matches the region that shows high iron line-to-continuum emission ratios (Fig.\ 2),
indicating that the enhanced iron emissions is mostly caused by an increase in
the iron abundance. The width of the arc is $\simeq1.2^\prime$, and the average
projected distance of the arc to the X-ray peak is about $2^{\prime}$
($33.6h_{70}^{-1}$ kpc). In the arc the abundance shows apparent spatial
variations ranging from 0.9 \solarZ\ to 1.3 \solarZ, while the abundance averaged
over the whole arc is $1.0\pm0.15$ \solarZ. There is also an asymmetric
high-abundance region located close to the X-ray peak, where the iron
line-to-continuum intensity ratio is high (Fig.\ 2), too. In the innermost
regions and the regions show high abundances, the typical abundance errors
are about $20-30$\%. In the outer regions the errors increase to about $50-100$\%.

To crosscheck our results further we extracted the spectra from 6 circular or pie
regions as are defined in Figure 3c and fitted them with the same spectral model
as is used above. We find that the results (Table 1) are generally consistent with
those implied in the 2-dimensional temperature and abundance maps. If we lower the
abundances of the arc (regions B and C) or the central region (region A), the fits
become significantly worse and unacceptable. By studying the two-dimensional
fit-statistic contours of temperature and iron abundance at the 68\%, 90\% and 99\%
confidence levels for the high-abundance arc (region B+C) and its surrounding
regions (regions D and F) we are confident that the iron abundance in the arc region
is higher than those of their neighboring regions at a significance of 90\% (Fig.\ 4a).

\section{Discussion and Conclusion}
We find that there is a remarkable high-abundance arc region at about
$2^{\prime}$ ($33.6h_{70}^{-1}$ kpc) from the X-ray peak of HCG 62, which spans
from south to northwest. The average abundance in the arc is higher than that in
its neighboring regions by a factor of about 2. A part of the high-abundance arc
roughly coincides with the outer edge of the southwest X-ray cavity, while with
the current data no other particular structure such as a cold front can be
identified as associated with the arc. Nor is there significant gas pressure
variations across the arc. This arc may be caused by the project effect of a part
of an optically thin spherical shell that has been excessively metal-enriched.
By studying its geometry (Fig. 5) we estimate that the part shell may has an
inclination angle (the angle between the axis of the shell and line of sight)
of 40 degrees, a thickness of $d=16h_{70}^{-1}$ kpc, a base radius of
$a=31.2h_{70}^{-1}$ kpc, and a radius of $R=36h_{70}^{-1}$ kpc. The iron mass
involved in such a part shell is calculated to be
$6.0\times 10^{6}h_{70}^{-2.5}$ \solarM.
It is also possible that the geometry is a spherical cap (Fig. 5) with roughly
the same inclination angle, whose height, base radius and radius are
$h=25h_{70}^{-1}$ kpc, $a=40.3h_{70}^{-1}$ kpc and $R=45h_{70}^{-1}$ kpc,
respectively. In this case the calculated iron mass is
$6.5\times 10^{6}h_{70}^{-2.5}$ \solarM.
Alternatively, if the high-abundance
arc is a ring with a cross-section radius of $8.3h_{70}^{-1}$ kpc and a radius
of $23h_{70}^{-1}$ kpc, the calculated iron mass is reduced to
$2.8\times 10^{6}h_{70}^{-2.5}$ \solarM.
Note that without further observational constraints it is difficult to figure
out the actual 3-dimensional geometry of the high-abundance region. However,
since the calculated iron mass is proportional to the square root of the assumed
volume, the bias of the estimated iron mass should be relatively small. The
involved iron mass shall have a conservative lower limit of about
$3\times 10^{6}h_{70}^{-2.5}$ \solarM,
which accounts for about 3\% of the iron that has been synthesized in NGC 4778
assuming an age of 10 Gyr.

The off-center, asymmetric iron concentrations also have been revealed in a few
other systems with a large scatter in the deduced iron mass involved in the
high-abundance regions. In a \Chandra\ observation of the Perseus Cluster
Sanders et al.\ (2005) reported a high-abundance shell at about 93 kpc from the
central nucleus, which is located at the edge of a radio mini-halo with two
H$_{\alpha}$ filaments pointing towards it. The authors argued that this shell
may be a remnant of a metal-enriched bubble lifting from the cluster core. Based
on their data we estimate that the iron mass in the high-abundance shell is
$^{<}_{\sim}10^{6}$ \solarM. In the relaxed cluster Abell 1060 Hayakawa et al.\ (2006)
found that there is a high-abundance flat ``blob'' northeast of the central bright
galaxy. The iron mass deduced with the \XMM\ dataset is $1.9\times 10^{7}$ \solarM.
This value, however, is larger than that obtained by Hayakawa et al.\ (2004) with
the \Chandra\ observation by about one order of magnitude, which was ascribed to
the relatively low effective area of \Chandra\ for the Fe-K photons. In the rich
cluster Abell 2199 Johnstone et al.\ (2002) found that the abundance is higher in
a north-northeast region about $0.6^\prime$ ($22h_{70}^{-1}$ kpc) from the center.
In the poor cD cluster AWM 7 Furusho et al.\ (2003) showed that the spatial
distribution of the hardness ratio of the counts in $6-7$ keV to that in $2-6$ keV
infers the existence of two high-abundance structures where the abundance has the
peak values. The two structures are located at about 7 kpc northeast and southwest
of the clusters center, containing about $2\times10^{4}$ \solarM\ of iron. In these
systems it is estimated that the mass of iron in the high-abundance regions account
for about 0.02\% to 4\% of the total mass of iron that has been synthesized in the
past.

The off-center high abundance values may be interpreted as a central abundance
dip that can be caused by the resonance scattering (e.g., Gil'fanov et al.\ 1987;
Matsushita et al.\ 2003). However, by modeling the resonance scattering in the
Centaurus cluster and Abell 2199 Sanders and Fabian (2006b) found that the
degradation effect of the resonance scattering on the Fe-L and Fe-K lines are
only about 10\% and 30\%, respectively, which is insufficient to cause the
observed central abundance dip. Similarly, in HCG 62 the estimated column
density of the hot gas is 3.85$\times$10$^{20}$ cm$^{-2}$ in the direction of
of the group center, which is too small to weaken the intensities of the Fe-L
lines with large oscillator strengths in the line of sight. Thus the contribution
of the resonance scattering should be negligible.

By modeling the spectra extracted from the high-abundance arc, we obtained
the average metal abundances
$Z_{\rm Fe}$=$0.76_{-0.13}^{+0.15}$ \solarZ,
$Z_{\rm Mg}$=$0.10_{-0.10}^{+0.17}$ \solarZ,
$Z_{\rm Si}$=$0.40_{-0.19}^{+0.22}$ \solarZ\ and
$Z_{\rm S}$=$0.43_{-0.28}^{+0.29}$ \solarZ\ (90\% confidence level).
After comparing the measured metal abundance ratios with the theoretical
supernova synthesis models (the W7 model and the weighted SN II model in
Nomoto et al.\ 1997), we find that the SNe Ia contribute about
$91_{-22}^{+9}\%$ (90\% confidence level) of the iron in the high-abundance
arc. For the innermost $0.4^\prime$ ($6.7h_{70}^{-1}$
kpc) region the obtained abundances are
$Z_{\rm Fe}$=$0.46\pm0.07$ \solarZ,
$Z_{\rm Mg}$=$0.18_{-0.08}^{+0.10}$ \solarZ,
$Z_{\rm Si}$=$0.29_{-0.07}^{+0.08}$ \solarZ\ and
$Z_{\rm S}$=$0.30_{-0.14}^{+0.02}$ \solarZ,
which infers a rather consistent SN Ia contribution fraction of
$86_{-16}^{+11}\%$ for iron. We notice that the high SN Ia contribution
fractions in the high-abundance regions in HCG 62 are very close to those found
in the central regions of clusters and groups (Wang et al.\ 2005 and references
therein). Moreover, in HCG 62 we find that the iron mass-to-optical light ratio
is significantly higher in the high-abundance arc than in other regions (Fig. 4b).
So it is natural to speculate that the high-abundance gas may have been enriched
in the group's core region and then been moved outwards to the present position.

The high-abundance arc could have been formed during the previous episodes of AGN
activities that might also create the two X-ray cavities, since a part of the arc
roughly coincides with the outer edge of the southwest cavity, inferring that the
metal-enriched gas was pushed outwards as the cavity expanded. However, we notice
that there are several facts that cannot be explained naturally with the AGN scenario.
First, the X-ray cavities are similar in size and luminosity, and have nearly the
same distances to the NGC 4778's nucleus. This strongly suggests symmetrical AGN
outbursts in the past, if the origin of the cavities is indeed related to the AGN.
However, the high-abundance arc and the gas emission excess (Fig. 1c) are only
detected in the southwest. No hint for the existence of similar structures in the
northeast direction is found within the \Chandra\ ACIS field, or in even outer
regions (out to about $10^{\prime}$) based on the 2-dimensional mappings with the
\ROSAT\ PSPC data. Second, a close examination shows that the high-abundance arc
spans over a region that is much larger and wider than the outer edge of the southwest
cavity (Fig. 2 and 3). Third, unlike in the Perseus Cluster (Sanders et al.\ 2005)
where a high-abundance shell is detected at the edge of a radio mini-halo, no such
direct radio evidence for recent AGN activities is found on the 1.4 GHz map of
HCG 62 at the 90\% confidence level. Forth, results of recent N-body simulation
works cast some doubts on the efficiency of metal transport via AGN activities.
For example, it is concerned that the buoyant bubbles and/or the energetic jets
alone may not be capable of transporting and diffusing materials of the inner
regions outwards with a sufficiently high efficiency (Churazov et al. 2001;
Vernaleo \& Reynolds 2005; Heath et al. 2006) to create a structure like the
perfectly shaped high-abundance arc seen in HCG 62. Finally, it should also be
noticed that multi-band observations show that there is no evidence for a currently
active AGN at the center of NGC 4778, and the AGN origin of the X-ray cavities
has been questioned by Morita et al. (2006) based on their thermal dynamic analysis
as well.

Still, the possibility of the AGN origin of the high-abundance arc cannot be
completely ruled out with the current data quality and simulation techniques,
despite of the uncertainties mentioned above. It is expected that some of these
uncertainties will be carefully inspected in the next-generation simulations
with $1024^{3}$ or $2048^{3}$ particles by fully taking into account additional
complexities such as the large-angle jet precession and gas turbulence. On
the other hand, it is worthy to investigate the possible impacts of other
mechanisms on metal transport and redistribution, primarily including the
supernova-driven winds, starburst-driven winds, and galaxy mergers and collisions.
By adopting a B-band luminosity of $1.6\times10^{10}$\solarLB~ (Hickson et al. 1989)
for NGC 4778 and recent supernova rates (e.g., Sharon et al. 2006),
we find that the dynamic power of the supernova events temporally averaged over
the past few Gyr is $9.6\times10^{40}$ erg s$^{-1}$. Obviously, the metal-enriched
gas in the high-abundance arc in HCG 62 could have been lifted to the present
location from inner regions by the supernova winds, if only $\sim$10\% of the
total supernova dynamic energy accumulated in the past 1 Gyr was used. However,
this mechanism has the difficulty to account for the asymmetric geometry of the
high-abundance arc, unless a rather anisotropic distribution of supernova winds
is reasonably introduced to create a major concentration of iron in the southwest
direction. A similar conclusion can be applied to the galactic winds driven by
recent or ongoing starbursts as well, for which no positive observational evidence
is found in literatures.

It is also possible that the high-abundance arc was formed in a recent merger event.
Since in a merger metals in the infalling galaxy is expected to form a stripe behind
it due to the ram-pressure stripping (Schindler et al. 2005), according to the
geometry of the high-abundance arc we suspect that most metals in the arc is likely
to originate in the central metal accumulation of the primary galaxy (NGC 4778).
Roettiger et al. (1997) showed that, if the velocity of the infalling galaxy exceeds
the speed of sound, a bow shock will form to inhibit the ram-pressure stripping of
the infalling gas, which helps a significant amount of the infalling gas penetrate
the core of the primary galaxy. A large fraction of the metal-enriched gas originally
accumulated in the inner regions of the primary galaxy is expected to be expelled
outwards in an anisotropic way during such a supersonic core passage. On the other
hand, if the speed of the infalling galaxy is subsonic, the infalling gas may have
been completely stripped before or during the core passage (Fabian \& Daines 1991).
In this case a metal outflow in the primary galaxy may also have the chance to arise
due to the momentum transfer from the infalling galaxy to the primary galaxy. The
impact of mergers on metal redistribution depends not only on the masses and velocities
of the colliding galaxies, but also on the collision angle, as well as other complicated
effects caused by, e.g., gas turbulence and magnetic field. So far these effects have
not been fully understood, and have not been fully accessed in the numerical studies,
but they are potentially crucial in the dense environment in and around the group's
core where the gas density is much higher than the gas density of the cluster
halo. In-depth numerical simulations taking into these effects are invoked to
inspect whether a high-abundance arc can be created in the hot gas during a merger
with the size and geometry comparable to what are observed in HCG 62.

On the observational aspect, there are indeed some hints for a recent minor merger
in HCG 62. First, the excess gas emission observed in $1.2^{\prime}-2^{\prime}$ in
the southwest of the X-ray peak, which roughly coincides with the inner edge of the
high-abundance arc, can be ascribed to an excess in gas density since there is no
particular temperature or abundance variations associated with this feature (\S3).
A recent minor merger may be a reasonable explanation for such a gas excess.
Second, the optical kinematic study of NGC 4778 by Spavone et al.\ (2006), who
used a long-slit positioned approximately in the east-west direction, revealed
both the twisting in the position angle of isophotes and the nuclear counterrotation
in the innermost region of the galaxy, apparently inferring that there is a
$\sim600$ pc core decoupled from the whole galaxy. The authors also showed that
the velocity dispersion distribution of the outer halo is strongly distorted in
the southwest direction, which cannot be interpreted naturally in terms of a
projection effect due to the presence of NGC 4776. Based on these peculiarities
the authors argued that the group must have suffered a recent minor merger. It
is possible that the same merger event triggered AGN activities to create the
low-density regions shown as the X-ray cavities, whose age is estimated to be
$\sim10$ Myr (Morita et al.\ 2006). However, the case may be more complicated,
since by using the X-ray and radio data we failed to detect any distinct evidence
for either the distortion of gas morphology, or the existence of a bow shock or
a front, which are expected to be visible on $\sim$ Gyr timescales since the
merger occurred (e.g., Roettiger et al. 1997).

By taking into account both the iron directly blown into the intra-group medium
during the SN Ia explosions and the iron lost in the stellar winds, we estimate
the enrichment times (B\"{o}hringer et al.\ 2004) using the method that was
described in details in Wang et al.\ (2005). We find that the times needed to
enrich the high-abundance arc and the innermost region are $1.0\pm0.1$ Gyr and
$0.2\pm0.1$ Gyr, respectively. If metals in the innermost region does have been
accumulating continuously since the destruction of the original central metal
accumulation, we estimate that the average speed of the metal outflow is
$\sim100$ km s$^{-1}$. Thus, the energy needed to blast the high-abundance gas
outwards is estimated to be $\sim3\times10^{56}$ erg.

\section{Summary}
We identify a remarkable high-abundance arc region at about $2^{\prime}$
($33.6h_{70}^{-1}$ kpc) from the center of NGC 4778, the dominating galaxy of
HCG 62. In the arc the average abundance is higher than that in the neighboring
regions by a factor of about 2 with the measured metal abundance ratios nicely
consistent with the dominance of the SN Ia yields. The arc contains
$>3\times 10^{6}h_{70}^{-2.5}$ \solarM\ of iron, or about $>3$\% of the cumulative
mass of iron that has been synthesized in the galaxy, which is rather high comparing
with the clusters of galaxies that exhibit similar high-abundance structures. A
part of the high-abundance arc roughly coincides with the outer edge of the southwest
X-ray cavity. With current data quality it is difficult to determine the origin of
this high-abundance arc. The optical photometric and kinematic evidence suggests
that the structure is likely to be formed in a recent merger, whereas the AGN origin
still cannot be completely ruled out.

\acknowledgments
We thank Yipeng Jing, Marilena Spavone and the referee of this paper for their
valuable suggestions and comments. This work was supported by the National Science
Foundation of China (Grant No. 10273009, 10233040 and 10503008), and Shanghai Key
Projects in Basic Research No. 04JC14079.

\clearpage
\begin{deluxetable}{ccccccccc}

  \tabletypesize{\scriptsize}
    \tablecolumns{9}
    \tablewidth{0pc}
    \tablecaption{Best-fit Spectral Models}
    \tablehead
    {
      \multicolumn{4}{c}{Projected}&&\multicolumn{3}{c}{Deprojected}&notes\\
      \cline{2-4}\cline{6-8}
      \colhead{Region No.}&\colhead{kT(keV)}&\colhead{$Z$(\solarZ)}&$\chi^2/{\rm dof}$&
      &\colhead{kT(keV)}&\colhead{$Z$(\solarZ)}&$\chi^2/{\rm dof}$\\
      \hline
      \multicolumn{8}{c}{Annular Regions\tablenotemark{a}}
    }
    \startdata

    1&$0.75\pm0.01$&$0.53_{-0.06}^{+0.12}$&143.6/117&&
    $0.71\pm0.01$&$0.86_{-0.27}^{+0.36}$&1412.1/1213\\

    2&$0.74\pm0.01$&$0.34_{-0.07}^{+0.03}$&223.4.5/138&&
    $0.81\pm0.01$&$0.41^{+0.09}_{-0.06}$&\nodata\\

    &$0.85_{-0.02}^{+0.05}$&$0.37\pm0.07$&71.1/64&&\nodata&\nodata&\nodata&without
    cavities\\

    &$0.91_{-0.04}^{+0.04}$&$0.37_{-0.09}^{+0.12}$&67.3/57&&\nodata&\nodata&\nodata&NE
    cavity only\\

    &$0.78_{-0.02}^{+0.02}$&$0.38_{-0.09}^{+0.13}$&48.9/53&&\nodata&\nodata&\nodata&SW
    cavity only\\

    3&$1.03_{-0.02}^{+0.01}$&$0.37_{-0.04}^{+0.07}$&37.38/46&&
    $1.00\pm0.02$&$0.46\pm0.09$&\nodata\\

    4&$1.18_{-0.05}^{+0.03}$&$0.47_{-0.09}^{+0.10}$&174.5.2/163&&
    $0.99_{-0.06}^{+0.04}$&$0.46_{-0.19}^{+0.31}$&\nodata\\

    5&$1.31_{-0.03}^{+0.02}$&$0.58_{-0.09}^{+0.10}$&147.1.2/138&&
    $1.34_{-0.06}^{+0.10}$&$1.42_{-0.49}^{+1.24}$&\nodata\\

    6&$1.31\pm0.04$&$0.34\pm0.07$&140.14/150&&
    $1.52_{-0.22}^{+0.18}$&$1.03_{-0.67}^{+1.27}$&\nodata\\
    \cutinhead{Selected Circular and Pie Region\tablenotemark{b}}
    A&$0.74_{-0.01}^{+0.01}$&$0.53_{-0.07}^{+0.10}$&143.6/111&&\nodata&\nodata&\nodata\\
    B&$1.29_{-0.04}^{+0.03}$&$0.83_{-0.22}^{+0.30}$&53.3/46&&\nodata&\nodata&\nodata\\
    C&$1.23_{-0.03}^{+0.04}$&$0.87_{-0.16}^{+0.38}$&72.5/46&&\nodata&\nodata&\nodata\\
    B+C&$1.28_{-0.03}^{+0.02}$&$0.84_{-0.15}^{+0.19}$&68.1/46&&\nodata&\nodata&\nodata\\
    D&$0.97_{-0.03}^{+0.02}$&$0.37_{-0.07}^{+0.10}$&67.6/46&&\nodata&\nodata&\nodata\\
    E&$0.81_{-0.01}^{+0.01}$&$0.25_{-0.04}^{+0.03}$&77.6/46&&\nodata&\nodata&\nodata\\
    F&$1.20_{-0.14}^{+0.10}$&$0.17_{-0.16}^{+0.10}$&74.6/94&&\nodata&\nodata&\nodata\\
    \enddata
    \tablenotetext{a}{Spectra were extracted in six annular regions (No.$1-6$) spanning
    over 0\arcmin-0.44\arcmin, 0.44\arcmin-0.88\arcmin, 0.88\arcmin-1.33\arcmin,
    1.33\arcmin-1.77\arcmin, 1.77\arcmin-2.21\arcmin\ and 2.21\arcmin-2.65\arcmin, respectively.
    For the second annulus, we also fitted the spectrum extracted when the two X-ray cavities
    were excluded and the spectra of the northeast and southwest cavities, respectively.
    Errors are quoted at the 90\% confidence level.}
    \tablenotetext{b}{Spectra were extracted in the circular and pie regions defined
    in Figure 3c.
    }
\end{deluxetable}

\clearpage
  \begin{figure}
    \begin{center}
      \includegraphics[width=.8\textwidth]{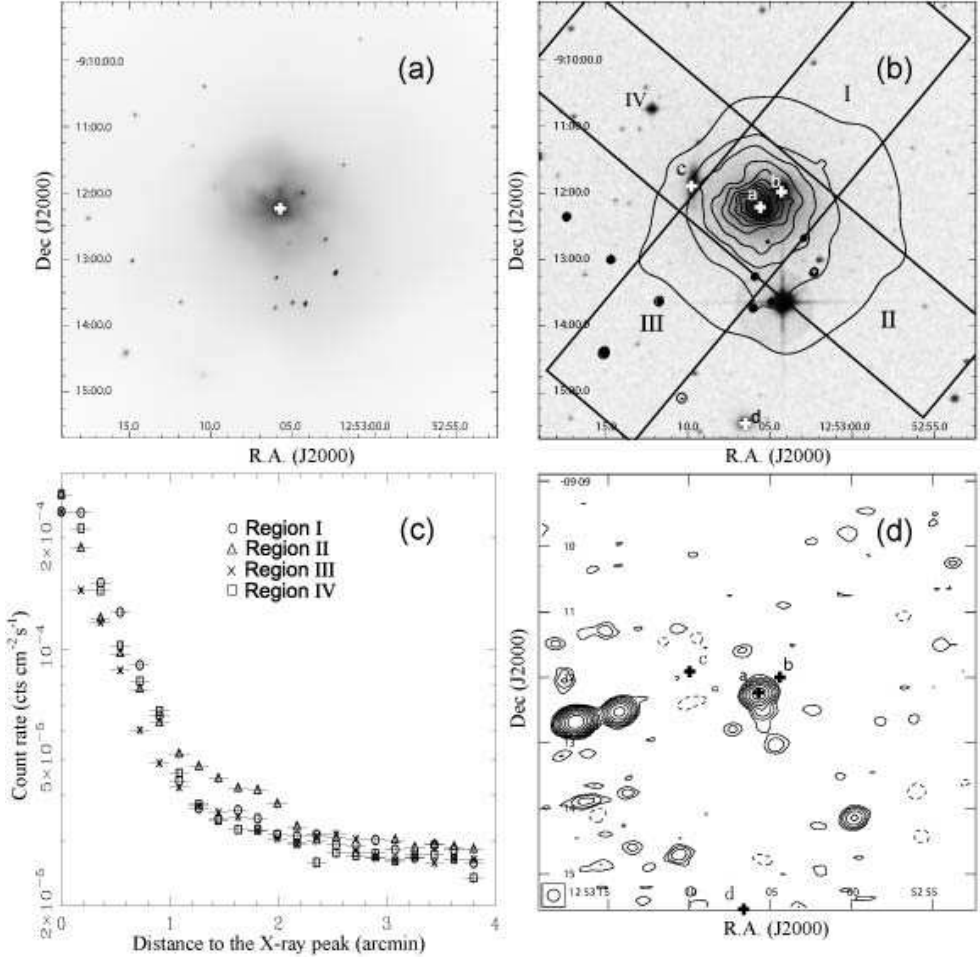}
      \caption{(a): \Chandra\ ACIS S3 image of the central
    $6.6^{\prime}\times6.6^{\prime}$ region of HCG 62 in 0.7--7 keV, which
    is plotted in the square-root scale. The image has been
    exposure-corrected, and smoothed by using the CIAO tool csmooth with a minimum
    significance of 3 and a maximum significance of 5. The optical center of the
    dominant galaxy NGC 4778 is marked as a `+'.
    (b): DSS optical image for the same sky field on which the X-ray contours
    are overlaid. Four bright member galaxies NGC 4778, NGC 4776, NGC 4761
    and NGC 4764 are marked as a, b, c and d, respectively. The strip regions
    are defined to extract the X-ray surface brightness profiles shown in (c).
    (c): X-ray surface brightness profiles extracted along the strips I, II,
    III and IV, along with the errors quoted at the 68\% confidence level.
    (d): VLA 1.43 GHz radio map for nearly the same field of view as in (a) and (b).
    Four bright members are also marked as in (b).
    }
    \end{center}
  \end{figure}

  \begin{figure}
    \begin{center}
      \includegraphics[width=0.8\textwidth]{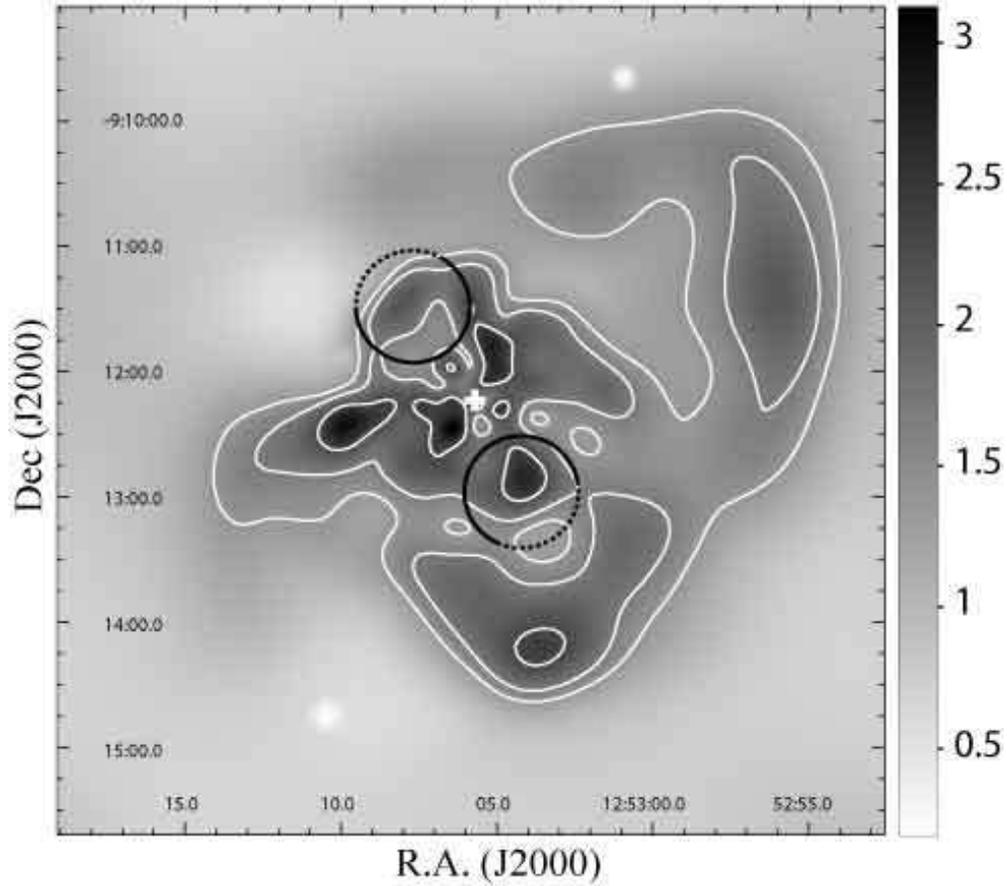}
      \caption{Distribution of the exposure-corrected and background-subtracted
      hardness ratio of the emission in 0.85--1.2 keV to that in 0.3--0.65 keV
      and 1.4--5 keV for the central $6.6^{\prime}\times6.6^{\prime}$ region,
      on which the ratio contours are overlaid with levels 1.4, 1.7 and 2.3. The
      image has been smoothed in the same way as in Figure 1a. The X-ray peak is
      marked with a `+', and the locations of the X-ray cavities are marked with
      black circles, on which the dotted part indicates where the cavity boundary
      is poorly constrained.
      }
    \end{center}
  \end{figure}

  \begin{figure}
    \begin{center}
      \includegraphics[width=0.8\textwidth]{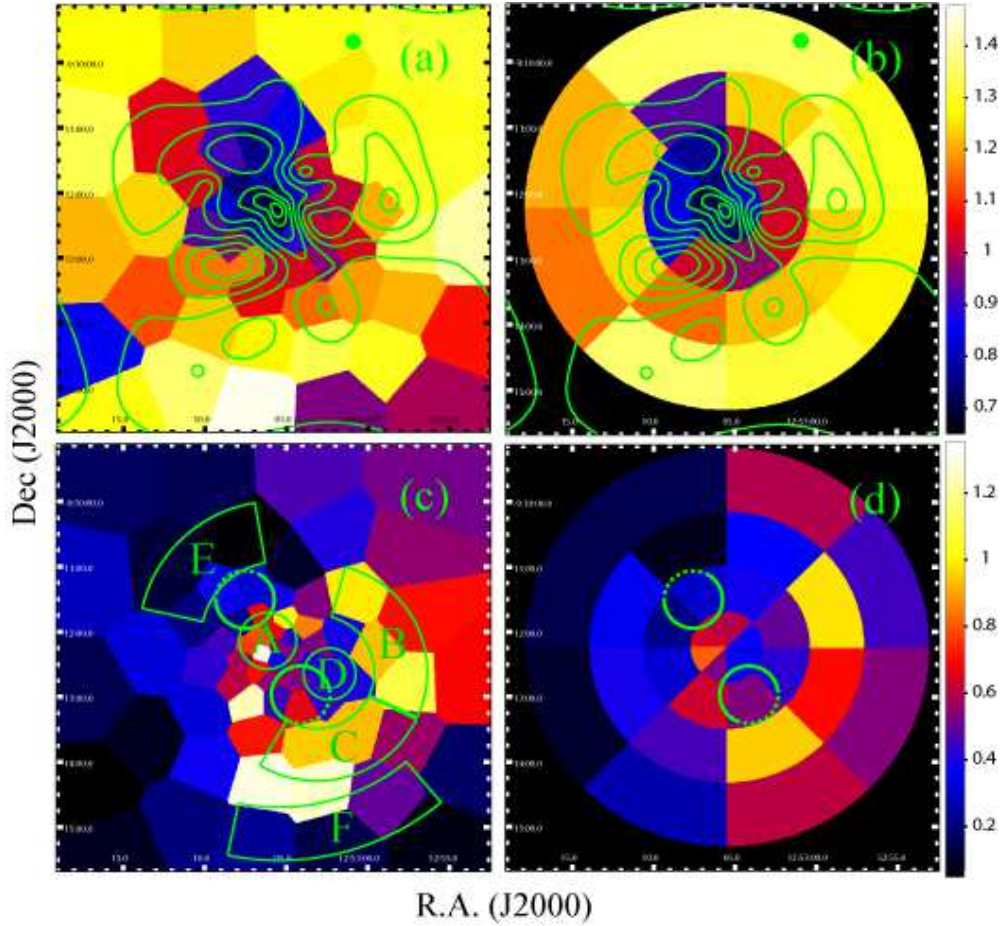}
      \caption{
    (a)-(b): Temperature maps for the central $6.6^{\prime}\times6.6^{\prime}$ region
    calculated by using the CVT binning algorithm (a) and pie binning algorithm (b), on
    which the linear contours of the exposure-corrected hardness ratio of the emission
    in 1.4--5 keV to that in 0.3--0.65 keV, which are both continuum-dominated so that
    the hardness ratio can be used to trace the temperature gradients, are overlaid.
    (c)-(d): Abundance maps for the central $6.6^{\prime}\times6.6^{\prime}$ region
    calculated by using the CVT binning algorithm (c) and pie binning algorithm (d), on
    which the X-ray cavities are marked with green circles (see Fig. 2). In (c) the
    circular and pie regions (A--F) are defined to extract the spectra for model fittings
    (\S3.2.3).
    }
    \end{center}
  \end{figure}

  \begin{figure}
    \begin{center}
      \includegraphics[height=0.75\textwidth]{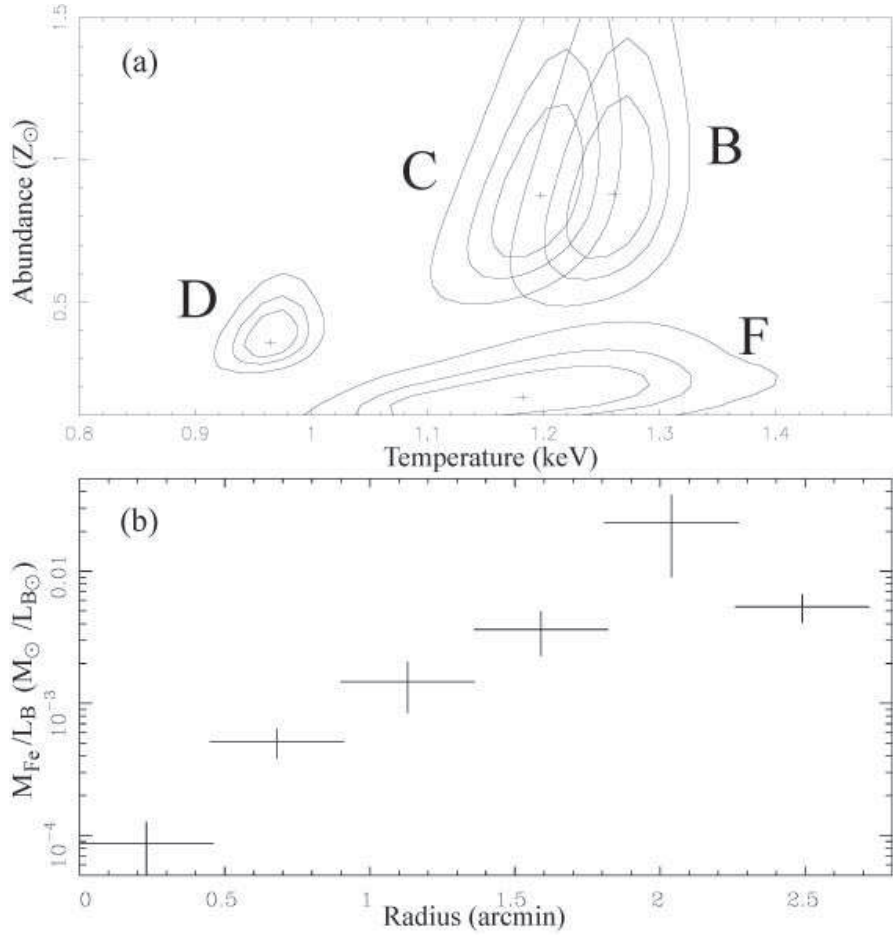}
      \caption{
    (a): Two-dimensional fit-statistic contours of temperature and abundance at the
    68\%, 90\% and 99\% confidence levels for regions B, C, D and F as are defined
    in Figure 3c.
    (b): Averaged radial distribution of the ratio of iron mass to B-band luminosity.
      \label{fig_stat_cont_01}}
    \end{center}
  \end{figure}

  \begin{figure}
    \begin{center}
      \includegraphics[height=0.35\textwidth]{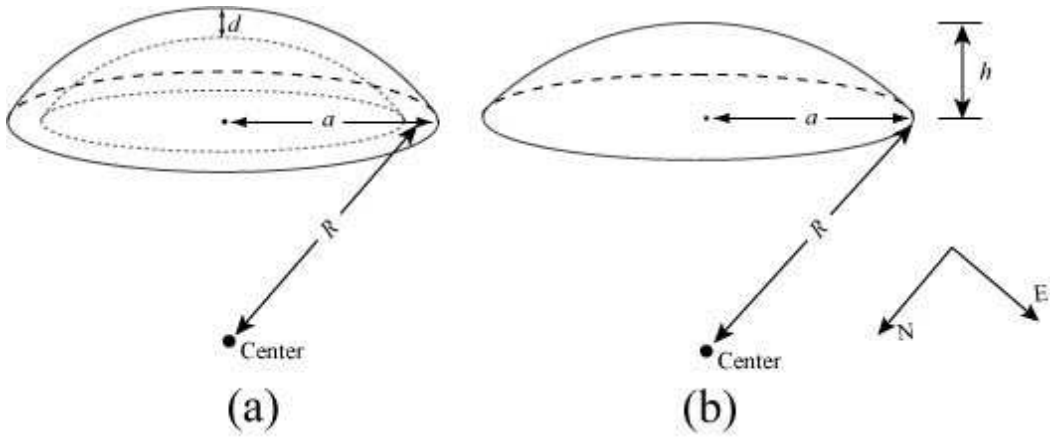}
      \caption{
    Assumed geometries for the high-abundance arc region. (a): a part shell.
    (b): a spherical cap. In the figures $a$, $R$, $d$ and $h$ are the base radius,
    radius, thickness and height, respectively. The center of HGC 4778 is marked with
    a black dot at the bottom.
      \label{fig_stat_cont_02}}
    \end{center}
  \end{figure}


\end{document}